\documentclass{article}

\usepackage{spconf}

\usepackage{microtype}
\microtypesetup{nopatch=footnote}

\usepackage{mathtools,amssymb,bm,empheq,nicefrac,upgreek,siunitx}

\usepackage{array,graphicx,makecell,nicematrix}
\usepackage[caption=false,labelformat=parens,subrefformat=parens]{subfig}

\usepackage{cite}
\usepackage{xcolor}
\usepackage[
    bookmarks=true,
    bookmarksnumbered=true,
    colorlinks=true,
    allcolors=blue,
    urlcolor=black
]{hyperref}
\usepackage[nameinlink,capitalise]{cleveref}
\usepackage{orcidlink}

\usepackage[subtle,title=normal,sections=normal,mathdisplays=normal,mathspacing=normal,lists=tight]{savetrees}

\let\oldcite\cite
\renewcommand{\cite}[1]{\textcolor{blue}{\oldcite{#1}}}

\expandafter\def\expandafter\normalequation\expandafter{\equation}
\expandafter\def\expandafter\endnormalequation\expandafter{\endequation}
\renewenvironment{equation}{\abovedisplayskip=3pt\belowdisplayskip=3pt\normalequation}{\endnormalequation}
\AtBeginEnvironment{align}
{%
  \setlength{\abovedisplayskip}{3pt plus 1pt minus 2pt}%
  \setlength{\belowdisplayskip}{3pt plus 1pt minus 2pt}%
  \setlength{\abovedisplayshortskip}{3pt plus 1pt minus 2pt}%
  \setlength{\belowdisplayshortskip}{3pt plus 1pt minus 2pt}%
  \setlength{\jot}{3pt}%
}

\begin{document}
\title{Frequency-Weighted Soft-Constrained Spatially Selective Active Noise Control for Open-Fitting Hearables}
\name{
Tong Xiao$^{1}$,
Reinhild Roden$^{2}$,
Matthias Blau$^{2}$,
and Simon Doclo$^{1}$
\thanks{This research was funded by the Deutsche Forschungsgemeinschaft
(DFG, German Research Foundation) -- Project-ID 352015383 -- SFB 1330 C1,
and under Germany's Excellence Strategy -- EXC 2177/2 -- Project ID 390895286.}
}
\address{
$^{1}$ Department of Medical Physics and Acoustics and Cluster of Excellence Hearing4all,\\
Carl von Ossietzky Universit\"at Oldenburg, Germany\\
$^{2}$ Institut f\"ur H\"ortechnik und Audiologie,
Jade Hochschule, Oldenburg, Germany
}

\ninept
\maketitle

\begin{abstract}
Recently, soft-constrained spatially selective active noise control (SSANC) for open-fitting hearables has been proposed to enable a flexible trade-off between noise reduction and desired-speech preservation using a scalar trade-off parameter across all frequencies. This paper generalizes soft-constrained SSANC by introducing frequency-dependent weighting of the desired-speech preservation error while maintaining low-delay time-domain active control. Frequency-dependent weightings based on the speech intelligibility index (SII), the modified intermediate reference system (MIRS), the long-term average speech spectrum (LTASS), and a signal-dependent oracle are compared with scalar weighting using measured acoustic impulse responses in a multi-speaker scenario. The results show that LTASS and oracle weightings provide the largest speech-intelligibility improvements and comparable speech-quality improvements at substantially lower distortion levels than those obtained with SII and MIRS weightings. While the oracle weighting provides the best overall perceptual trade-off, LTASS weighting provides a comparable perceptual trade-off without requiring the clean desired-speech signal.
\end{abstract}


\section{Introduction}
\label{sec:introduction}

Speech communication is often degraded in multi-speaker environments, where a listener aims to follow a desired speaker in the presence of interfering speakers and other undesired acoustic sources. Hearables can enhance communication by preserving desired speech while attenuating undesired sounds at the listener's ear\cite{doclo2015,Chang2016Listening,gupta2022augmented}. Such devices typically comprise multiple outer microphones, an inner microphone, and a loudspeaker, hereafter referred to as the secondary source. Open-fitting hearables reduce the occlusion effect and improve physical comfort, but also lead to acoustic leakage of both desired and undesired sounds to the eardrum. 

To selectively reduce undesired sources while preserving the desired-speech source, spatial filtering can be applied to the outer-microphone signals. In the following, the term \emph{noise source} is used to denote any undesired source, including interfering speakers. Common spatial filtering approaches are minimum-variance-distortionless-response (MVDR) and minimum-power-distortionless-response (MPDR) beamforming\cite{VanTrees2002optimum,gannot2017consolidated}. These beamformers do not explicitly account for acoustic leakage or the secondary path from the secondary source to the inner microphone. In contrast, conventional active noise control (ANC) directly minimizes the inner-microphone signal power\cite{benois2022optimization,Iotov2025speech}, but attenuates both desired speech and noise. Spatially selective active noise control (SSANC) combines spatial filtering and active control to selectively attenuate noise while preserving desired speech at the inner microphone\cite{xiao2023spatial,xiao2024icassp,xiao2025fa,xiao2026iwaenc}. 

\begin{figure}[t]
    \centering
    \includegraphics[width=0.99\columnwidth]{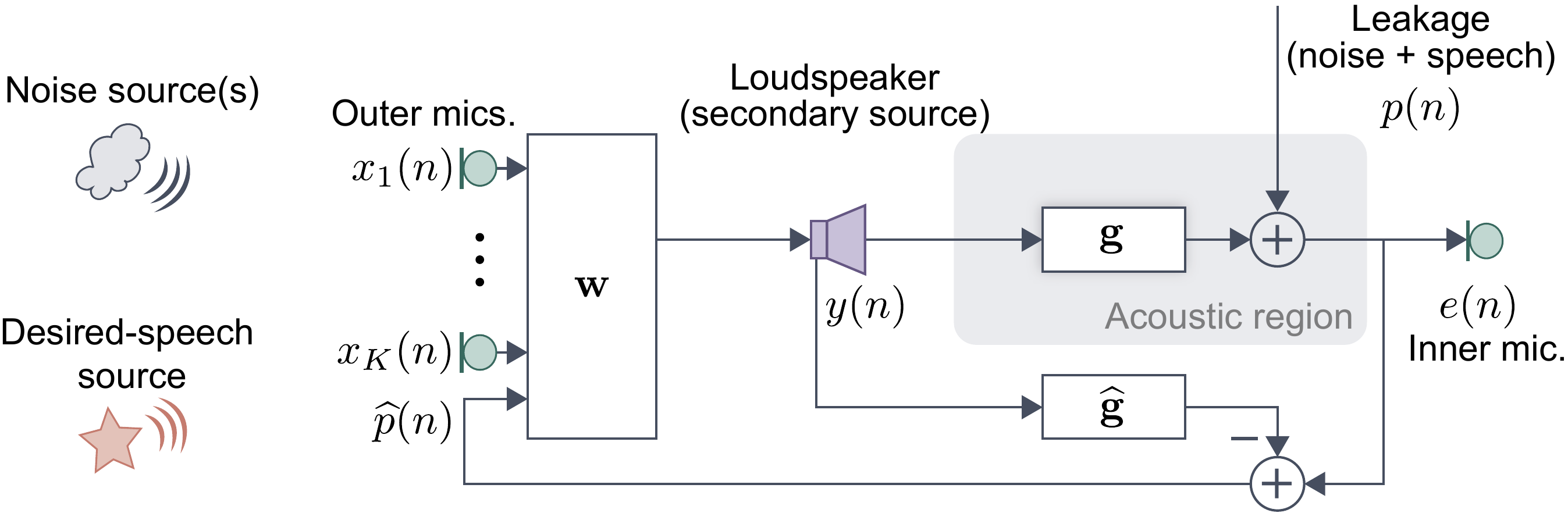}
    \vspace{-12pt}
    \caption{Block diagram of the SSANC system using $K$ outer-microphone signals $x_k(n)$, $k=1,\ldots,K$, together with the leakage estimate $\widehat{p}(n)$ computed from the inner-microphone signal $e(n)$, the secondary-source signal $y(n)$, and the secondary-path estimate $\widehat{\mathbf{g}}$. The control filter $\mathbf{w}$ generates $y(n)$, which is filtered by the secondary path $\mathbf{g}$ and acoustically combined with the leakage $p(n)$ to form $e(n)$.}
    \label{fig:block_diagram_ssanc}
    \vspace{-8pt}
 \end{figure}

Hard-constrained SSANC imposes desired-speech preservation exactly, which may restrict achievable noise reduction\cite{xiao2023spatial,xiao2024icassp,xiao2025fa}, whereas soft-constrained SSANC relaxes this requirement using a quadratic penalty with a scalar trade-off parameter\cite{xiao2025soft}. In multichannel speech enhancement, the speech-distortion-weighted multichannel Wiener filter (SDW-MWF) uses a scalar trade-off parameter to balance noise reduction and speech distortion\cite{Doclo2002gsvd}. Later approaches introduced frequency-dependent weighting to control this trade-off using speech-presence probabilities (SPP)\cite{Ngo2009spp}, psychoacoustic masking\cite{Defraene2012psychoacoustic}, and perceptual frequency weighting, e.g., based on the speech intelligibility index (SII)\cite{Kodrasi2015curvature}. These frequency-dependent approaches are typically implemented using block-based processing, which may be incompatible with the low processing delay required in SSANC.

In this paper, we generalize soft-constrained SSANC using frequency-dependent weighting of the desired-speech preservation error, while maintaining low-delay time-domain active control. We compare frequency-dependent weightings based on SII, the modified intermediate reference system (MIRS), the long-term average speech spectrum (LTASS), and a signal-dependent oracle, with scalar weighting as a baseline. The results show that LTASS and oracle weightings provide the largest speech-intelligibility improvements and comparable speech-quality improvements at substantially lower distortion levels than those obtained with SII and MIRS weightings. While the oracle weighting provides the best overall perceptual trade-off, LTASS weighting provides a comparable perceptual trade-off without requiring the clean desired-speech signal.

\section{Signal Model}
\label{sec:signal_model}

We consider a hearable comprising $K$ outer microphones, one inner microphone, and one secondary source as shown in \cref{fig:block_diagram_ssanc}. The outer-microphone signals are denoted by $x_k(n)$, $k=1,\ldots,K$, with $n$ denoting the discrete time index. Since acoustic feedback from the secondary source to the outer microphones can be effectively cancelled using established feedback-cancellation techniques\cite{spriet2005adaptive,schepker2019null}, the outer-microphone signals are modeled as containing only the external sound field, i.e.,
\begin{equation}
    x_k(n)=x_{k,s}(n)+x_{k,v}(n), \qquad k=1,\ldots,K,
    \label{eq:x_k}
\end{equation}
where $(\cdot)_s$ and $(\cdot)_v$ denote the desired-speech and noise components, respectively. Similarly, the inner-microphone signal $e(n)$ and the leakage signal $p(n)$ at the inner microphone can be decomposed as
\begin{equation}
    e(n)=e_s(n)+e_v(n), \qquad p(n)=p_s(n)+p_v(n).
    \label{eq:e_and_p}
\end{equation}

The inner-microphone signal $e(n)$ is formed by the acoustic superposition of the leakage $p(n)$ and the secondary-source signal $y(n)$ filtered by the secondary path $\mathbf{g}$ from the secondary source to the inner microphone, i.e.,
\begin{equation}
    e(n)=p(n)+\mathbf{g}^{\mathcal{T}}\mathbf{y}(n),
    \label{eq:e_pGy}
\end{equation}
where $\mathbf{g}=[g_0\,\dots\,g_{L_g-1}]^\mathcal{T}$ and $\mathbf{y}(n)=[y(n)\,\dots\,y(n-L_g+1)]^\mathcal{T}$ denote the secondary-path filter and the secondary-source signal vector, respectively, with $L_g$ denoting the length of the secondary-path filter. $(\cdot)^\mathcal{T}$ denotes the transpose operator. We assume that an estimate $\widehat{\mathbf{g}}$ of the secondary path $\mathbf{g}$ is available.

SSANC employs a hybrid architecture that uses the outer-microphone signals together with a leakage estimate as inputs to the control filter. From~\labelcref{eq:e_pGy}, the leakage estimate is obtained as
\begin{equation}
    \widehat{p}(n)=e(n)-\widehat{\mathbf{g}}^\mathcal{T}\mathbf{y}(n).
    \label{eq:p_hat}
\end{equation}
We assume an exact secondary-path estimate, i.e., $\widehat{\mathbf{g}}=\mathbf{g}$, as commonly done in analytical studies of feedback and hybrid ANC systems\cite{Kuo1996active,Elliott2000}, such that $\widehat{p}(n)=p(n)$. In the following, we omit the hat on $\mathbf{g}$.

The secondary-source contribution at the inner microphone $\mathbf{g}^\mathcal{T}\mathbf{y}(n)$ in~\labelcref{eq:e_pGy} can be written as $(\mathbf{G}\mathbf{w})^\mathcal{T}\mathbf{x}(n)$, with the stacked control filter $\mathbf{w}$ defined as
\begin{equation}
    \mathbf{w}=[\mathbf{w}_1^\mathcal{T}\,\dots\,\mathbf{w}_K^\mathcal{T}\,\mathbf{w}_{K+1}^\mathcal{T}]^\mathcal{T}\in\mathbb{R}^{(K+1)L_w},
    \label{eq:w_time}
\end{equation}
where $\mathbf{w}_k=[w_{k,0}\,\dots\,w_{k,L_w-1}]^\mathcal{T}\in\mathbb{R}^{L_w}$, and $L_w$ denotes the control-filter length for each input channel. The block-diagonal convolution matrix $\mathbf{G}$ associated with the secondary path is defined as
\begin{subequations}
\begin{align}
    \mathbf{G}&=\mathrm{blkdiag}(\mathsf{G}\ldots\mathsf{G})\in\mathbb{R}^{(K+1)L\times(K+1)L_w},
    \label{eq:G_multi}
    \\[-2pt]
    \mathsf{G}&=
    \begin{bmatrix}
        g_0       & \cdots & 0      \\[-2pt]
        \vdots    & \ddots & \vdots \\[-6pt]
        g_{L_g-1} & \ddots & g_0    \\[-2pt]
        \vdots    & \ddots & \vdots \\[-0pt]
        0         & \cdots & g_{L_g-1}
    \end{bmatrix}
    \in\mathbb{R}^{L\times L_w},
    \label{eq:G}
\end{align}
\end{subequations}
with $L=L_g+L_w-1$.
The stacked input vector $\mathbf{x}(n)$ is defined as
\begin{equation}
    \mathbf{x}(n)=[\mathbf{x}_1^\mathcal{T}(n)\,\dots\,\mathbf{x}_K^\mathcal{T}(n)\,\widehat{\mathbf{p}}^\mathcal{T}(n)]^\mathcal{T}\in\mathbb{R}^{(K+1)L},
    \label{eq:x_time}
\end{equation}
where $\mathbf{x}_k(n)=[x_k(n)\,\dots\,x_k(n-L+1)]^\mathcal{T}\in\mathbb{R}^{L}$, and $\widehat{\mathbf{p}}(n)=[\widehat{p}(n)\,\dots\,\widehat{p}(n-L+1)]^\mathcal{T}\in\mathbb{R}^{L}$.
The leakage estimate can be extracted from $\mathbf{x}(n)$ as $\widehat{p}(n)=\mathbf{q}^\mathcal{T}\mathbf{x}(n)$, with
\begin{equation}
    \mathbf{q}=[\mathbf{0}^\mathcal{T}\ldots\mathbf{0}^\mathcal{T}\bm{\updelta}^\mathcal{T}]^\mathcal{T}\in\mathbb{R}^{(K+1)L},
    \;\;
    \bm{\updelta}=[1\,0\,\dots\,0\,]^\mathcal{T}\in\mathbb{R}^{L}.
\end{equation}
Therefore, the inner-microphone signal in~\labelcref{eq:e_pGy} can be written as
\begin{equation}
    \boxed{e(n)=(\mathbf{q}+\mathbf{G}\mathbf{w})^\mathcal{T}\mathbf{x}(n)  } 
    \label{eq:e_qGwx}
\end{equation}

\section{Soft-Constrained Spatially Selective Active Noise Control}
\label{sec:soft_constrained_ssanc}

\subsection{Hard- and frequency-independent soft-constrained SSANC}
\label{sec:hard_and_frequency_independent_soft_ssanc}

SSANC aims to preserve the desired-speech component at the inner microphone as
\begin{equation}
    e_s(n)= x_{\mathrm{ref},s}(n-\Delta),
    \label{eq:es_xrefs_time}
\end{equation}
where $x_{\mathrm{ref},s}(n)$ denotes the desired-speech component at a reference microphone, and $\Delta$ is a nonnegative desired-speech delay. As in\cite{xiao2023spatial,xiao2024icassp,xiao2025fa}, this requirement can be written in matrix form as
\begin{equation}
    \mathbf{H}(\mathbf{q}+\mathbf{G}\mathbf{w})=\bm{\updelta}_{\Delta},
    \label{eq:ssanc_constraint_time}
\end{equation}
where
\begin{subequations}
\begin{align}
    \mathbf{H}&=[\mathbf{H}_1\,\dots\,\mathbf{H}_{K+1}]
    \in\mathbb{R}^{(L_a+L_h+L-1)\times(K+1)L},
    \label{eq:H_time}
    \\
    \bm{\updelta}_{\Delta}&=
    [\,\underbrace{0\,\dots\,0}_{L_a}\,
    \underbrace{0\,\dots\,0}_{\Delta}\,
    \underbrace{1\;\;0\,\dots\,0}_{L_h+L-1-\Delta}\,]^\mathcal{T}
    \in\mathbb{R}^{L_a+L_h+L-1}.
    \label{eq:delta_Delta}
\end{align}
\end{subequations}
Here, $\mathbf{H}_k\in\mathbb{R}^{(L_a+L_h+L-1)\times L}$ denotes the convolution matrix associated with the relative impulse response (ReIR) $\mathbf{h}_k=[h_{k,-L_a}\, \allowbreak \dots \allowbreak \, h_{k,L_h-1}]^\mathcal{T}$ between the reference microphone and the $k$-th microphone, where $L_a$ and $L_h$ denote the lengths of the acausal and causal parts of $\mathbf{h}_k$, respectively. $\mathbf{H}_k$ has the same structure as $\mathbf{G}$ in~\labelcref{eq:G}.

The desired-speech preservation requirement in~\labelcref{eq:ssanc_constraint_time} can be imposed exactly using hard-constrained SSANC\cite{xiao2023spatial,xiao2024icassp,xiao2025fa}, formulated as
\begin{equation}
    \min_{\mathbf{w}} \; \mathcal{E}\{e^2(n)\}+\beta\|\mathbf{w}\|_2^2
    \qquad \text{s.t.} \;
    \mathbf{H}(\mathbf{q}+\mathbf{G}\mathbf{w})=\bm{\updelta}_{\Delta},
    \label{eq:ssanc_optimization_hard}
\end{equation}
where $\mathcal{E}\{\cdot\}$ denotes the expectation operator, $\|\cdot\|_2$ denotes the $\ell_2$ norm, and $\beta$ is a nonnegative regularization parameter. 

To provide a controllable trade-off between desired-speech preservation and noise reduction, frequency-independent soft-constrained SSANC\cite{xiao2025soft} relaxes the hard constraint by penalizing the desired-speech preservation error. The optimization problem is formulated as
\begin{equation}
    \min_{\mathbf{w}}
    \mathcal{E}\{e^2(n)\}+\beta\|\mathbf{w}\|_2^2
    +\mu\left\|\mathbf{H}(\mathbf{q}+\mathbf{G}\mathbf{w})
    -\bm{\updelta}_{\Delta}\right\|_2^2,
    \label{eq:ssanc_optimization_soft_scalar}
\end{equation}
where $\mu$ is a nonnegative scalar trade-off parameter. Increasing $\mu$ places greater emphasis on desired-speech preservation, whereas decreasing $\mu$ provides more freedom for noise reduction. For $\mu=0$, the soft-constrained formulation reduces to conventional ANC, whereas for $\mu\rightarrow\infty$ it approaches the hard-constrained SSANC solution\cite{xiao2025soft}.

\subsection{Proposed frequency-weighted soft-constrained SSANC}
\label{sec:frequency_weighted_soft_ssanc}

To generalize soft-constrained SSANC to frequency-dependent weighting while maintaining low-delay time-domain active control, we apply a frequency-weighting filter to the desired-speech preservation error in the controller optimization. This preserves sample-by-sample time-domain processing without requiring block-based processing. The weighting filter is defined as $\mathbf{f}=[f_0\,\dots\,f_{L_f-1}]^\mathcal{T}\in\mathbb{R}^{L_f}$, where $L_f$ denotes the weighting-filter length. The corresponding convolution matrix $\mathbf{F}$ is constructed from the weighting-filter coefficients with the same structure as $\mathsf{G}$ in~\labelcref{eq:G}, and has dimensions $\mathbf{F}\in\mathbb{R}^{(L_f+L_a+L_h+L-2)\times(L_a+L_h+L-1)}$. The proposed frequency-weighted soft-constrained SSANC optimization problem is formulated as
\begin{equation}
    \boxed{
    \min_{\mathbf{w}}
    \mathcal{E}\{e^2(n)\}+\beta\|\mathbf{w}\|_2^2 
    +\mu\left\|\mathbf{F}\left[
    \mathbf{H}(\mathbf{q}+\mathbf{G}\mathbf{w})
    \!-\!\bm{\updelta}_{\Delta}\right]\right\|_2^2  } \!\!
\label{eq:ssanc_optimization_soft_weighted}
\end{equation}
Compared with the frequency-independent formulation in~\labelcref{eq:ssanc_optimization_soft_scalar}, the weighting filter $\mathbf{f}$ allows the penalty on the desired-speech preservation error to be weighted differently across frequency, while $\mu$ controls the overall trade-off between desired-speech preservation and noise reduction. Frequency regions emphasized by the weighting filter contribute more strongly to the preservation-error penalty, whereas less emphasized regions provide more freedom for noise reduction.

The solution of~\labelcref{eq:ssanc_optimization_soft_weighted} can be obtained as
\begin{align}
    \mathbf{w}
    =-\big(\bm{\Phi}_{\mathbf{rr}}
    &+\mu\mathbf{G}^\mathcal{T}\mathbf{H}^\mathcal{T}
    \bm{\Phi}_f\mathbf{H}\mathbf{G}\big)^{-1}
    \notag
    \\[-0pt]
    \times\big[\bm{\upphi}
    &+\mu\mathbf{G}^\mathcal{T}\mathbf{H}^\mathcal{T}
    \bm{\Phi}_f\left(\mathbf{H}\mathbf{q}
    -\bm{\updelta}_{\Delta}\right)\big],
    \label{eq:w_soft_weighted}
\end{align}
where
\vspace{-6pt}
\begin{subequations}
\label{eq:soft_weighted_matrix_definitions}
\begin{alignat}{2}
    \!\! \bm{\Phi}_{\mathbf{xx}} &= \mathcal{E}\{\mathbf{x}(n)\mathbf{x}^\mathcal{T}(n)\} &&\in\mathbb{R}^{(K+1)L\times(K+1)L}, \label{eq:Phi_xx}  \\[-3pt]
    \!\! \bm{\Phi}_{\mathbf{rr}} &= \mathbf{G}^\mathcal{T}\bm{\Phi}_{\mathbf{xx}}\mathbf{G}+\beta\mathbf{I} &&\in\mathbb{R}^{(K+1)L_w\times(K+1)L_w}, \label{eq:Phi_rr}\\[-3pt]
    \!\! \bm{\upphi} &= \mathbf{G}^\mathcal{T}\bm{\Phi}_{\mathbf{xx}}\mathbf{q} &&\in\mathbb{R}^{(K+1)L_w}, \label{eq:phi}  \\[-3pt]
    \!\! \bm{\Phi}_f &= \mathbf{F}^\mathcal{T}\mathbf{F} &&\in\mathbb{R}^{(L_a+L_h+L-1)\times(L_a+L_h+L-1)}, \!\!\! \label{eq:Phi_f} 
\end{alignat}
\end{subequations}
with $\mathbf{I}$ denoting the identity matrix.

\begin{figure}[t]
    \centering
    \includegraphics[width=0.65\columnwidth]{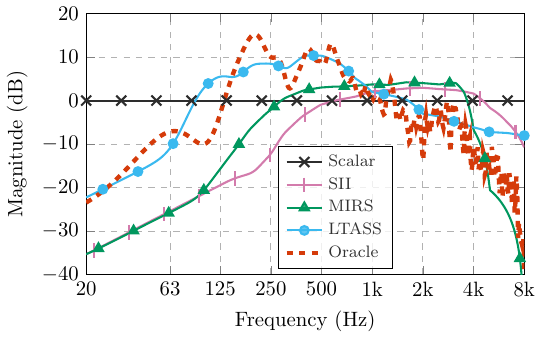}
    \vspace{-13pt}
    \caption{Magnitude responses of the scalar, SII, MIRS, LTASS, and oracle weighting filters for an example speech signal at a sampling frequency of \qty{16}{\kilo\hertz}. All weighting filters are normalized to have unit mean-square magnitude response.}
    \label{fig:speech_weighting_filters}
    \vspace{-4pt}
\end{figure}

\subsection{Speech weighting filters}
\label{sec:speech_weighting_filters}

Let $F(\omega)$ denote the frequency response of the weighting filter $\mathbf{f}$, where $\omega$ denotes the normalized radian frequency. For a fair comparison, all weighting filters are normalized such that
\begin{equation}
    \frac{1}{N_\omega}\sum\nolimits_{m=1}^{N_\omega}|F(\omega_m)|^2=1,
    \label{eq:weighting_normalization}
\end{equation}
where $\{\omega_m\}_{m=1}^{N_\omega}$ denotes the normalized radian-frequency grid. The frequency-dependent weighting filters were implemented as $L_f$-tap finite impulse response (FIR) filters using the MATLAB \texttt{fir2} function, which uses the frequency-sampling method. Their normalized magnitude responses are shown in~\cref{fig:speech_weighting_filters}. The following five weightings are considered.
\begin{itemize}
    \item \textit{Scalar:} This weighting corresponds to the frequency-independent formulation in~\labelcref{eq:ssanc_optimization_soft_scalar} and applies uniform weighting across frequency. One possible choice is $\mathbf{f}=[1\,\,0\,\,\dots\,\,0]^\mathcal{T}$. 
    \item \textit{SII:} This weighting uses the band-importance coefficients defined in the speech intelligibility index standard\cite{ANSI1997} directly as the weighting-filter magnitude before normalization, emphasizing frequency regions that contribute more strongly to speech intelligibility.
    \item \textit{MIRS:} This weighting is based on the modified intermediate reference system receive characteristic specified in ITU-T Recommendation P.830\cite{ITU_P830}, emphasizing the frequency range relevant for speech communication.
    \item \textit{LTASS:} This weighting is based on the long-term average speech spectrum reported in\cite{Byrne1994}, assigning larger weights to frequency regions containing more speech energy on average.
    \item \textit{Oracle:} This signal-dependent weighting is based on the power spectral density (PSD) $P_{\mathrm{ref},s}(\omega)$ of the desired-speech component $x_{\mathrm{ref},s}(n)$ at the reference microphone, with $|F(\omega)| \propto \sqrt{P_{\mathrm{ref},s}(\omega)}$. It represents the case where the spectral distribution of the complete desired-speech component is known.
\end{itemize}
%
%

\section{Evaluation}
\label{sec:evaluation}

\subsection{Acoustic setup}
\label{sec:acoustic_setup}

\begin{figure}[t]
    \centering
    \captionsetup[subfloat]{captionskip=0pt,farskip=0pt}
    \subfloat[]{\includegraphics[height=0.37\linewidth]{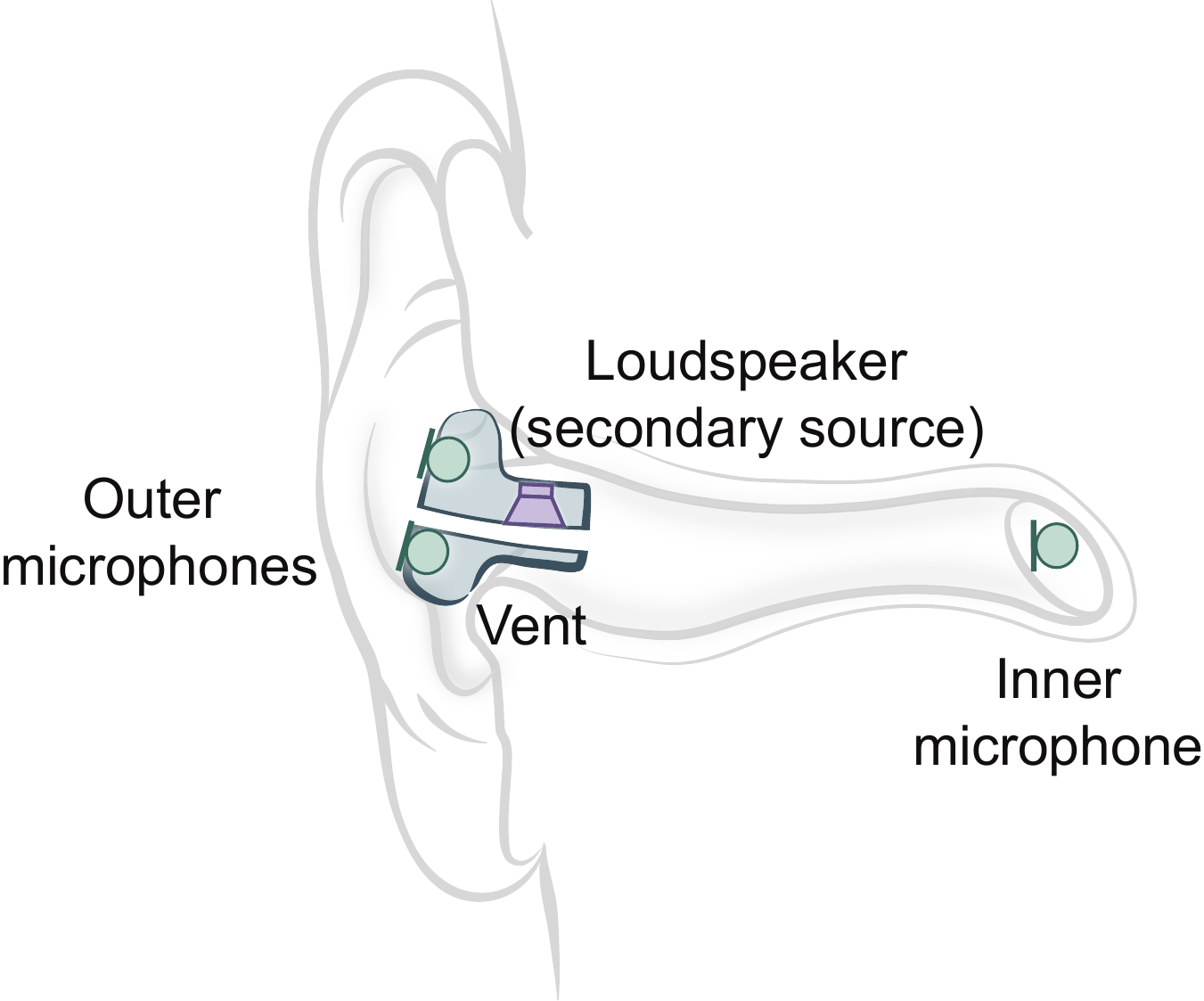}  \label{fig:Ear}} 
    \subfloat[]{\includegraphics[height=0.37\linewidth]{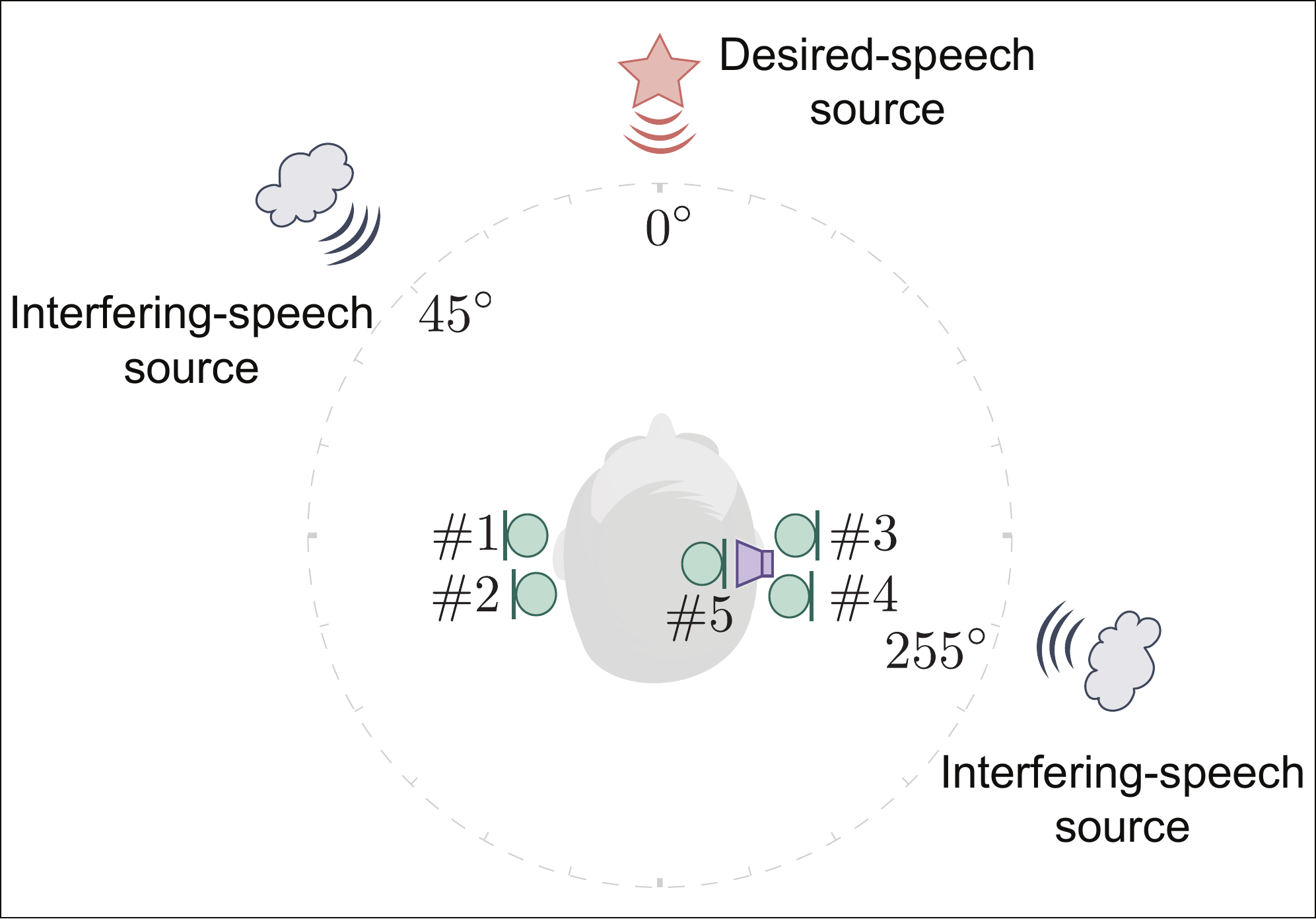}  \label{fig:Scene}}
    \vspace{-10pt}
    \caption{(a) Open-fitting hearable comprising multiple outer microphones, one inner microphone, and a loudspeaker as the secondary source. Acoustic leakage reaches the inner microphone through the vent and around the hearable. (b) Acoustic setup comprising four outer microphones (\#1--\#4), one inner microphone (\#5), and a secondary source at the right hearable. The desired-speech source is located at $0^\circ$, and two interfering speech sources are located at $45^\circ$ and $255^\circ$.}
    \label{fig:Ear_and_Scene}
    \vspace{-9pt}
\end{figure}

For the evaluation, we considered a pair of open-fitting hearables inserted into both ears of a GRAS 45BB-12 KEMAR Head \& Torso simulator, as illustrated in~\cref{fig:Ear_and_Scene}. Four outer microphones (\#1--\#4) and one inner microphone at the right eardrum (\#5) were considered. The outer driver of the right hearable was used as the secondary source, and the right entrance microphone \#3 was selected as the reference microphone.

The acoustic paths were based on measured impulse responses obtained in a moderately reverberant room with dimensions $7\times6\times2.7$~\qty{}{\meter} and a reverberation time of $T_{60}\approx\qty{330}{\milli\second}$. The desired-speech source at $0^\circ$ and two interfering speech sources at $45^\circ$ and $255^\circ$ were positioned approximately \qty{1.8}{\meter} from the center of the KEMAR. The microphone signals were generated by convolving each source signal with the corresponding measured impulse responses. The desired-source ReIRs were obtained by least-squares estimation from white-noise microphone signals generated using the measured impulse responses for the $0^\circ$ direction, with microphone \#3 as reference. The measured impulse response between the secondary source and the inner microphone was used as the secondary path. The ReIRs and secondary path were assumed to be accurately known to isolate the algorithmic performance from parameter-estimation errors. The source levels were adjusted to obtain an inner-microphone leakage SNR of \qty{-5}{\decibel}, with equal interference powers. For the leakage-SNR analysis, the interference level was varied to obtain additional SNRs of \qty{-10}{\decibel}, \qty{0}{\decibel}, and \qty{5}{\decibel}.

The desired-speech signal was constructed by concatenating utterances \texttt{005-007} from speaker \texttt{p361} of the VCTK dataset\cite{veaux2017cstr}. The interfering speech sources at $45^\circ$ and $255^\circ$ consisted of utterances \texttt{018-020} from speaker \texttt{p243} and utterances \texttt{316-318} from speaker \texttt{p238}, respectively. All signals had a duration of \qty{15}{\second} and were sampled at \qty{16}{\kilo\hertz}.

The desired-speech delay was fixed at $\Delta=64$ samples (\qty{4}{\milli\second}). The filter lengths were set to $L_w=2000$, $L_g=300$, $L_a=L_h=3000$, and $L_f=1025$. The regularization parameter was set to $\beta=\lambda_{\max}(\mathbf{G}^\mathcal{T}\bm{\Phi}_{\mathbf{xx}}\mathbf{G})/10^9$, where $\lambda_{\max}(\cdot)$ denotes the largest eigenvalue. To investigate the influence of frequency weighting, the trade-off parameter was varied as $\mu\in[10^{-8},10^4]$.

\subsection{Evaluation metrics}
\label{sec:evaluation_metrics}

The performance is evaluated in terms of noise reduction (NR), broadband SNR improvement, desired-speech distortion, speech quality, and intelligibility. The NR and broadband SNR improvement are defined as
\begin{subequations}
\label{eq:evaluation_nr_snr}
    \begin{align}
        \mathrm{NR} &= 10\log_{10}\sum\nolimits_n p_v^2(n)-10\log_{10}\sum\nolimits_n e_v^2(n), \label{eq:evaluation_nr}\\
        \Delta\mathrm{SNR} &= 10\log_{10}\frac{\sum\nolimits_n e_s^2(n)}{\sum\nolimits_n e_v^2(n)}-10\log_{10}\frac{\sum\nolimits_n p_s^2(n)}{\sum\nolimits_n p_v^2(n)}. \label{eq:evaluation_snr}
    \end{align}
\end{subequations}

The amount of desired-speech distortion is quantified using the scale-invariant signal-to-distortion ratio (SI-SDR)~\cite{LeRoux2019sdr}, with the optimal scaling factor $\eta={\sum_n e_s(n)x_{\mathrm{ref},s}(n-\Delta)}/{\sum_n x_{\mathrm{ref},s}^2(n-\Delta)}$. The SI-SDR is then defined as
\begin{equation}
    \mathrm{SI\mbox{-}SDR}=10\log_{10}\frac{\sum\nolimits_n[\eta x_{\mathrm{ref},s}(n-\Delta)]^2}{\sum\nolimits_n[e_s(n)-\eta x_{\mathrm{ref},s}(n-\Delta)]^2}.
    \label{eq:evaluation_sisdr}
\end{equation}

Speech quality and intelligibility are evaluated using narrowband perceptual evaluation of speech quality (PESQ)~\cite{Rix2001PESQ} and extended short-time objective intelligibility (ESTOI)~\cite{Jensen2016algorithm}, respectively, with $x_{\mathrm{ref},s}(n-\Delta)$ as the reference signal. The improvements $\Delta$PESQ and $\Delta$ESTOI are computed as the corresponding metric differences between $e(n)$ and $p(n)$.


\subsection{Results}
\label{sec:results}

\begin{figure}
    \centering
    \includegraphics[width=0.99\columnwidth]{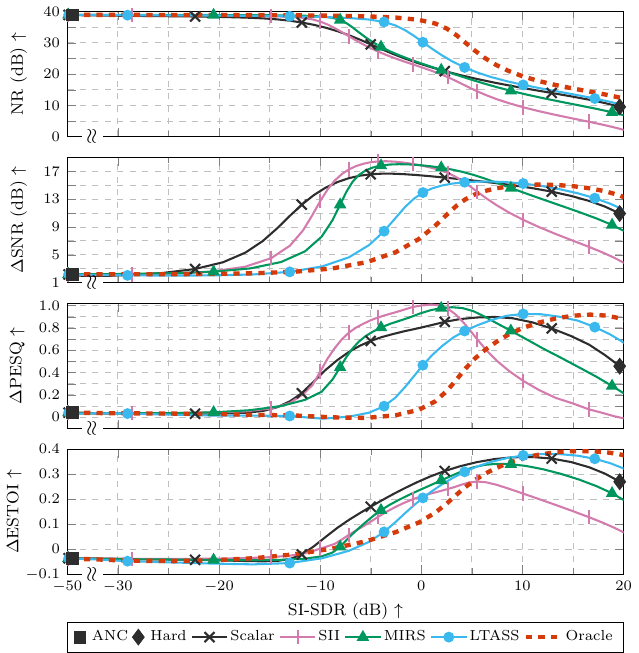}
    \vspace{-10pt}
    \caption{Performance trade-offs obtained by varying the trade-off parameter $\mu$ for the five weightings. NR, $\Delta$SNR, $\Delta$PESQ, and $\Delta$ESTOI are shown as functions of SI-SDR, with conventional ANC and hard-constrained SSANC included as reference operating points. Since the results are nearly identical in the SI-SDR range from $-50$ to $-30$~dB, this range is compressed for visualization.}
    \label{fig:results_time_2sources_soft}
    \vspace{-6pt}
\end{figure}

\Cref{fig:results_time_2sources_soft} shows NR, $\Delta$SNR, $\Delta$PESQ, and $\Delta$ESTOI as functions of SI-SDR for the scalar, SII, MIRS, LTASS, and oracle weightings. For each weighting, $\mu$ is varied to obtain different operating points, illustrating the trade-offs between noise reduction, speech quality, intelligibility, and desired-speech distortion. Conventional ANC achieves the largest NR of \qty{39.0}{\decibel}, but strongly distorts the desired speech, whereas hard-constrained SSANC exhibits the opposite trade-off, achieving a high SI-SDR of \qty{19.6}{\decibel} but at an NR of \qty{9.5}{\decibel}.

For the scalar weighting, varying $\mu$ produces the expected transition from conventional ANC toward hard-constrained SSANC, consistent with the limiting behavior discussed in~\Cref{sec:hard_and_frequency_independent_soft_ssanc} and in~\cite{xiao2025soft}. Compared with scalar weighting, SII weighting increases the maximum $\Delta$SNR from \qty{16.7}{\decibel} to \qty{18.5}{\decibel} and the maximum $\Delta$PESQ from $0.90$ to $1.01$. However, its maximum $\Delta$PESQ occurs at an SI-SDR of \qty{1.0}{\decibel}, compared with \qty{7.2}{\decibel} for scalar weighting, indicating larger desired-speech distortion. Its maximum $\Delta$ESTOI is also lower, decreasing from $0.37$ for scalar weighting to $0.27$. MIRS shows similar but less pronounced behavior, achieving a maximum $\Delta$PESQ of $0.99$ at an SI-SDR of \qty{3.0}{\decibel}, with a maximum $\Delta$ESTOI of $0.34$.

In contrast, LTASS and oracle weightings achieve high perceptual performance at higher SI-SDR than scalar weighting. LTASS weighting achieves a maximum $\Delta$PESQ of $0.93$ at an SI-SDR of \qty{10.1}{\decibel}, while the oracle weighting achieves $0.92$ at \qty{16.7}{\decibel}. Both weightings also provide the largest $\Delta$ESTOI improvements of $0.38$ and $0.39$, respectively. Although LTASS and oracle weightings yield lower maximum $\Delta$SNR than the scalar, SII, and MIRS weightings, they achieve comparable speech-quality improvements and larger intelligibility improvements at substantially higher SI-SDR, demonstrating that larger broadband SNR improvement does not necessarily imply a more favorable perceptual trade-off. Although the oracle weighting provides the best overall perceptual trade-off, the practically applicable LTASS weighting achieves closely comparable performance.

To assess the influence of leakage SNR, \Cref{tab:snr_sweep_metrics} reports the maximum $\Delta$PESQ and $\Delta$ESTOI, together with the corresponding SI-SDR, for leakage SNRs from \qty{-10}{\decibel} to \qty{5}{\decibel}. Across all leakage SNRs, SII and MIRS weightings achieve larger maximum $\Delta$PESQ at relatively low SI-SDR, whereas LTASS and oracle weightings provide the largest $\Delta$ESTOI improvements and maintain high maximum $\Delta$PESQ at substantially higher SI-SDR.

\begin{table}[t]
\centering
\caption{Maximum $\Delta$PESQ and $\Delta$ESTOI over $\mu$ with corresponding SI-SDR (\unit{\decibel}) in parentheses for different leakage SNRs (\unit{\decibel}).}
\label{tab:snr_sweep_metrics}
\setlength{\tabcolsep}{2.4pt}
\begin{tabular}{c|ccccc}
\hline
SNR & Scalar & SII & MIRS & LTASS & Oracle \\
\hline
\multicolumn{6}{c}{$\Delta$PESQ} \\
\hline
$-10$ & 0.81 (5.3) & 0.95 (0.0) & 0.92 (1.8) & 0.85 (9.8) & 0.83 (14.7) \\
$-5$  & 0.90 (7.2) & 1.01 (1.0) & 0.99 (3.0) & 0.93 (10.1) & 0.92 (16.7) \\
$0$   & 0.60 (7.2) & 0.71 (1.8) & 0.68 (4.1) & 0.61 (11.7) & 0.63 (18.0) \\
$5$   & 0.34 (8.8) & 0.43 (2.6) & 0.42 (5.1) & 0.37 (15.0) & 0.41 (19.8) \\
\hline
\multicolumn{6}{c}{$\Delta$ESTOI} \\
\hline
$-10$ & 0.44 (8.9) & 0.34 (5.3) & 0.41 (7.2) & 0.45 (11.5) & 0.47 (14.7) \\
$-5$  & 0.37 (11.0) & 0.27 (5.5) & 0.34 (7.5) & 0.38 (11.8) & 0.39 (15.0) \\
$0$   & 0.27 (10.9) & 0.18 (5.6) & 0.25 (8.9) & 0.28 (13.6) & 0.30 (18.0) \\
$5$   & 0.17 (12.5) & 0.08 (6.5) & 0.15 (10.3) & 0.18 (16.7) & 0.20 (18.5) \\
\hline
\end{tabular}
\vspace{-8pt}
\end{table}

\section{Conclusion}
\label{sec:conclusion}

This paper generalized soft-constrained SSANC by introducing frequency-dependent weighting of the desired-speech preservation error while maintaining low-delay time-domain active control. Results show that SII and MIRS weightings achieve larger maximum SNR and speech-quality improvements than LTASS and oracle weightings, but at larger desired-speech distortion. In contrast, LTASS and oracle weightings provide the largest intelligibility improvements and comparable speech-quality improvements at substantially lower distortion than those obtained with SII and MIRS weightings. These trends remain consistent across the considered leakage SNRs. While the oracle weighting provides the best overall perceptual trade-off, the practically applicable LTASS weighting provides closely comparable performance without requiring the clean desired-speech signal. 

\bibliographystyle{IEEEtran}
\bibliography{main}

\end{document}